\documentclass[12pt,aps,reprint,groupedaddress,showpacs,nofootinbib,prd]{revtex4-1}
\usepackage{amsmath,amssymb,mathrsfs,graphicx,slashed,ragged2e}

\usepackage[bottom]{footmisc}
\usepackage{hyperref}

\newcommand{\mb}{\overbar{m}}

\newcommand{\bl}{\boldsymbol{\ell}}
\newcommand{\bn}{\boldsymbol{n}}

\newcommand{\overbar}[1]{\mkern 1.5mu\overline{\mkern-1.5mu#1\mkern-1.5mu}\mkern 1.5mu}

\begin{document}

\title{Gravitational perturbations of a Kerr black hole in $f(R)$ gravity}

\author{Arthur George Suvorov}
\email{arthur.suvorov@tat.uni-tuebingen.de}
\affiliation{Theoretical Astrophysics, IAAT, University of T{\"u}bingen, Germany}


\date{\today}

\begin{abstract}
 
Modified theories of gravity are often built such that they contain general relativity as a limiting case. This inclusion property implies that the Kerr metric is common to many families of theories. For example, all analytic $f(R)$ theories with vanishing constant term admit the Kerr solution. In any given theory, however, the response of the gravitational field to astrophysical disturbances is tied to the structure of the field equations. As such, even if black holes are Kerr, the underlying theory can, in principle, be probed through gravitational distortions. In this paper, we study linear perturbations of a Kerr black hole in $f(R)$ gravity using the Newman-Penrose formalism. We show that, as in general relativity, the equations governing the perturbed metric, which depend on the quadratic term of the function $f$, completely decouple.

\end{abstract}


\maketitle

\section{Introduction}

The Kerr metric describes the geometry surrounding an isolated, rotating black hole \cite{kerrmetric44}. Amongst the numerous special properties of this solution are a set of uniqueness results effectively stating that, in general relativity (GR), astrophysically stable black holes must be Kerr \cite{cart1,rob1,mars1,gurl15}. The implication is that a detection of prominent non-Kerr attributes (such as those described in Refs. \cite{papa1,papa2}) of such an object would strongly indicate a breakdown of GR in the strong field regime \cite{bhtests,bambi08}. Many experiments aiming to, in part, search for such features, such as those pertaining to black hole X-ray reflection spectroscopy \cite{kerrjoh16,kerre1,kerrspec2}, gravitational wave (GW) data analysis \cite{kerre2,kerre3X,gw17}, or direct imaging via black hole `shadows' \cite{bhshad2,bhshad3,m87shad}, have been carried out. To date, all relevant data are consistent with the Kerr metric, though there is room for alternative theories \cite{kerre3}.

Indeed, motivated by a number of theoretical and observational issues (e.g. the elusive nature of dark energy \cite{darken1,darkenergy}), an abundance of modified theories of gravity aiming to extend GR in one way or another have been introduced \cite{modgravths}. An aspect that is often demanded of these extended theories is that they reduce to GR in some appropriate limit, much in the same way that GR reduces to the Newtonian theory in the weak field regime \cite{will06}. As such, although unlikely to be unique, the Kerr metric is still in fact an exact solution in many cases \cite{kerrinmod}. In analytic $f(R)$ theories, the focus of this paper, the Kerr metric [or its asymptotically $\textrm{(anti-)}$de Sitter variant] is always a solution \cite{felice}. This implies that a validation of the Kerr metric does not necessarily favour GR amongst all possibilities \cite{morekerr}. However, since the field equations dictate the character of perturbations, the gravitational field surrounding a \emph{distorted} Kerr object, such as a black hole disturbed by an active accretion disk \cite{accrete} or a young remnant born out of a compact object merger or core collape \cite{kokqnm,ringing2}, will be different in non-GR theories \cite{morekerr,schfofr,pertkerrfr}. Future precision experiments in GW astronomy may be able to identify whether these sub-leading order signatures are present in the waveforms \cite{subtle}.

In GR, the properties of gravitational perturbations are encapsulated by the Teukolsky equations \cite{teu73,teu74}, which describe, using the Newman-Penrose (NP) language \cite{np62}, the dynamics of the perturbed Weyl scalars. The Weyl scalars can then be integrated to determine the energy and angular-momentum outfluxes at infinity \cite{sachs,teu73}. Moreover, as initially shown by Chrzanowski \cite{chr75,chrz76} (see also Refs. \cite{coh75,perttet,otherkerrmet}), the Weyl scalars actually contain enough information to fully reconstruct the metric of a (vacuum) perturbed Kerr spacetime in GR. Unfortunately, several obstacles prevent an immediate generalisation of these results to modified theories of gravity. For example, certain gauge choices, permissible in GR and necessary for the Chrzanowski procedure \cite{price07}, are inadmissible due to the existence of additional GW polarisation modes \cite{gwpols1,gwpols2}. Nevertheless, we show here that, in $f(R)$ theories of gravity, a set of decoupled, linear differential equations for the perturbed metric can still be derived. These can be solved to determine the distorted spacetime structure, which differs from its GR counterpart through an explicit dependence on the quadratic coefficient of the function $f$. 

This paper is organised as follows. In Sec. II we write down the $f(R)$ field equations, present a brief recap of the NP formalism, and give the standard Kinnersley null tetrad description of the Kerr metric \cite{kinn69}. In Section III we introduce metric perturbations and the $f(R)$ Teukolsky equations. In Section IV the decoupled metric equations, which constitute the main result, are derived. Some brief discussion is offered in Sec. V. 




\section{Field equations}

In this paper, we are interested in studying how perturbations of the Kerr metric behave in the $f(R)$ theory of gravity. In this class of theories, the Ricci scalar, $R$, in the Einstein-Hilbert Lagrangian is replaced by an arbitrary function of this quantity, $f(R)$. The vacuum field equations read\footnote{Throughout this paper, we adopt natural units with $G=c=1$, use a timelike $(+,-,-,-)$ metric signature, and denote complex conjugation by an overhead bar.} (see Ref. \cite{felice} for a review)
\begin{equation} \label{eq:fofr1}
0 = f'(R) R_{\mu \nu} -  \frac {f(R)} {2} g_{\mu \nu} + \left( g_{\mu \nu} \square - \nabla_{\mu} \nabla_{\nu} \right) f'(R),
\end{equation}
where $R_{\mu \nu} = R^{\alpha}_{\mu \alpha \nu}$ is the Ricci tensor, $g_{\mu \nu}$ is the metric tensor, and $\square = \nabla_{\mu} \nabla^{\mu}$ represents the d'Alembert operator. Setting $f(R) = R$ returns the vacuum Einstein equations. A simple rearrangement shows that the field equations \eqref{eq:fofr1} can also be written in a manner resembling the Einstein equations, viz.
\begin{equation} \label{eq:fofrfieldTEU}
\begin{aligned}
R_{\mu \nu}  =& \frac {1} {f'(R)} \Big[ \frac {f'(R)} {2} R g_{\mu \nu} +  g_{\mu \nu} \frac {f(R) - f'(R) R} {2} \\
&+\nabla_{\mu} \nabla_{\nu} f'(R) - g_{\mu \nu} \square f'(R) \Big] .
\end{aligned}
\end{equation}

Taking the trace of \eqref{eq:fofrfieldTEU} yields a constraint between the Ricci scalar and the function $f$,
\begin{equation} \label{eq:treqn}
3 \square f'(R) + R f'(R) -2 f(R) = 0.
\end{equation}


%


\subsection{Newman-Penrose variables and the Kerr metric}

Throughout this paper, we will make use of the well known NP formalism \cite{np62}. For our purposes, it is sufficient to note that this formalism introduces a set of null tetrads at each spacetime point, $\{\boldsymbol{\ell},\boldsymbol{n},\boldsymbol{m},\boldsymbol{\mb}\}$, which are used to decompose the metric (see below). Relevant curvature quantities, which depend on the metric and its derivatives, can thus be recast in terms of the tetrads and their derivatives, the latter of which are succinctly expressed through the twelve \emph{spin-coefficients}: $\kappa, \rho, \sigma, \tau, \lambda, \mu, \nu, \pi, \epsilon, \gamma, \beta$, and $\alpha$. These latter quantities are formed by applying the directional derivatives $D, \Delta, \delta$, and $\overbar{\delta}$, defined as
\begin{equation} \label{eq:directionalderivatives}
D \equiv \ell^{\mu} \nabla_{\mu}, \Delta \equiv n^{\mu} \nabla_{\mu}, \delta \equiv m^{\mu} \nabla_{\mu}, \overbar{\delta} \equiv \mb^{\mu} \nabla_{\mu},
\end{equation}
to the tetrad components, and are related to the Christoffel symbols \cite{np62}.

The various formulas of (pseudo-)Riemannian geometry are thus translated into relationships between the spin-coefficients, culminating in the construction of ten functions (called the Ricci-NP coefficients) which completely encode the components of the Ricci tensor ($\Phi_{ij}$ for $i,j=0,1,2$ and $\Lambda$), and five functions (Weyl scalars) which detail contractions of the Weyl tensor ($\psi_{i}$ for $i=0, \ldots, 4$). A major benefit of the approach is that, due to the compact nature of the formalism, the curvature quantities $\boldsymbol{\Phi}$ and $\boldsymbol{\psi}$ often enjoy a simple description when spacetime symmetries are imposed. For arbitrary (i.e. potentially non-Einstein) spacetimes, a complete description of the relationships between the NP symbols is given by Pirani \cite{pirani64}, which we will not repeat here.

Having briefly introduced the NP formalism, we now utilise it to describe the Kerr metric. To this end, we introduce the Kinnersley tetrad in Boyer-Lindquist coordinates $\{t,r,\theta,\phi\}$ \cite{kinn69},
\begin{equation} \label{eq:ellkerr}
\ell^{\mu} = \left( \frac {\left( r^2 + a^2 \right)} { r^2 - 2 M r + a^2 } , 1 ,0 \frac {a} {r^2 - 2 M r + a^2} \right),
\end{equation}
\begin{equation} \label{eq:nkerr}
n^{\mu} = \frac {\left( r^2 + a^2 , - r^2 + 2 M r -a^2, 0 ,a \right)} {2 \left( r^2 + a^2 \cos^2\theta \right)},
\end{equation}
and
\begin{equation} \label{eq:mkerr}
m^{\mu} = \frac { \left( i a \sin \theta, 0, 1, i \csc\theta \right) } {\sqrt{2} \left( r + i a \cos\theta \right)},
\end{equation}
where $M$ and $a$ represent the mass and the spin parameter of the black hole, respectively. From expressions \eqref{eq:ellkerr}--\eqref{eq:mkerr}, the components of the Kerr metric are given through the general formula
\begin{equation} \label{eq:kerrmet}
g_{\mu \nu} = \ell_{\mu} n_{\nu} + n_{\mu} \ell_{\nu} - m_{\mu} \mb_{\nu} - \mb_{\mu} m_{\nu}.
\end{equation}
For the Kerr metric, the non-zero NP spin-coefficients are given by \cite{teu73,np62}
\begin{equation} \label{eq:kerrnpquant}
\begin{aligned}
&\rho = -\left(r - i a \cos\theta \right)^{-1}, \,\,\, \beta = -\overbar{\rho} \cot\theta / 2 \sqrt{2}, \\
& \pi = i a \rho^2 \sin\theta / \sqrt{2}, \,\,\, \tau = - i a \rho \overbar{\rho} \sin\theta / \sqrt{2} , \\
&\mu = \rho^2 \overbar{\rho} \left(r^2 - 2 M r + a^2 \right) /2, \\
& \gamma = \mu + \rho \overbar{\rho} \left( r - M \right)/2, \,\,\, \alpha = \pi - \overbar{\beta}.
\end{aligned}
\end{equation}
Additionally, being a vacuum solution to GR, the Kerr metric has vanishing Ricci tensor, $R_{\mu \nu} = 0$, which implies that $\Phi_{ij} = \Lambda = 0$. The only non-zero Weyl scalar is $\psi_{2} = M \rho^3$ \cite{teu73}. As is evident from \eqref{eq:fofrfieldTEU}, the Kerr metric is a solution to all $f(R)$ theories which satisfy $f(0) = 0$ (i.e. those which have vanishing cosmological constant).

\section{Perturbations}

Through a slight abuse of notation, we begin by introducing a metric perturbation
\begin{equation} \label{eq:mettpt}
g_{\mu \nu} \rightarrow g_{\mu \nu} + h_{\mu \nu},
\end{equation}
where the metric $\boldsymbol{g}$ on the right-hand side of \eqref{eq:mettpt} is given by \eqref{eq:kerrmet}. Throughout this paper, perturbed quantities are denoted with a subscript $h$, while we omit the subscript $g$ on background quantities when no ambiguity arises. The mapping \eqref{eq:mettpt} similarly introduces perturbations into the null tetrads through \eqref{eq:kerrmet}, which therefore introduces perturbations into the spin-coefficients (e.g. $\rho \rightarrow \rho + \rho_{h}$), all of which are to be constrained through the perturbed version of the field equations \eqref{eq:fofrfieldTEU}. In Appendix A, we give the general (i.e. theory-independent) expressions for the perturbed NP quantities in terms of the perturbed metric \cite{lousto99}. We will make use of these expressions throughout the rest of this paper.


Moving forward, we will assume that the function $f$ is analytic so that it may be expanded as a Maclaurin series,
\begin{equation} \label{eq:maclaurin}
f(R) = R + \frac{a_{2}}  {2!} R^{2} + \cdots .
\end{equation}
Using \eqref{eq:maclaurin}, we have that, to leading order in $\boldsymbol{h}$, the perturbed trace equation \eqref{eq:treqn} reads
\begin{equation} \label{eq:perttrace}
3 a_{2} \square_{g} R_{h} - R_{h} = 0,
\end{equation}
which one may recognise as the Klein-Gordon equation over a Kerr background for $a_{2} \neq 0$. Equation \eqref{eq:perttrace}, which can be solved through a separation of variables \cite{brill72}, demonstrates that $f(R)$ gravity predicts the existence of a massive Ricci mode (sometimes called the scalaron) \cite{massgws,gong18}. One typically imposes\footnote{If we were to use the spacelike sign convention $(-,+,+,+)$ for the metric, this condition would read $a_{2} \geq 0$ instead.} $a_{2} \leq 0$ to avoid the so-called tachyonic instability \cite{bg11,starobinsky} (though cf. Ref. \cite{suv17}).




\subsection{Perturbed Ricci-NP coefficients}
As will prove useful, we proceed by computing various contractions of the perturbed Ricci tensor with the Kerr null tetrads. For example, noting that $R_{g} = 0$, we find that
\begin{equation} \label{eq:ricnp1}
\begin{aligned}
n_{\mu} \mb_{\nu} R_{h}^{\mu \nu}  &= \frac {n_{\mu} m_{\nu} \nabla^{\mu} \nabla^{\nu} f'(R_{h})} {f'(0)} \\
&= a_{2} n_{\mu} \mb_{\nu} \nabla^{\mu} \nabla^{\nu} R_{h} \\
&= a_{2} n_{\mu} \overbar{\delta} \nabla^{\mu} R_{h} \\
&= a_{2} \left[ \overbar{\delta} \left( n_{\mu} \nabla^{\mu} \right) - \left( \overbar{\delta} n_{\mu} \right) \nabla^{\mu} \right] R_{h} \\
&= a_{2} \left[ \overbar{\delta} \Delta - \overbar{\mu} \overbar{\delta} + \left( \alpha + \overbar{\beta} \right) \Delta \right] R_{h}.
\end{aligned}
\end{equation}
The first equality in \eqref{eq:ricnp1} holds as a consequence of orthogonality relationships between the background metric and the tetrad components (i.e. $n_{\mu} \mb_{\nu} g^{\mu \nu} = 0$), the second through the Maclaurin expansion \eqref{eq:maclaurin}, the next two from index rearrangement and a simple application of the Leibniz rule, respectively, while the final is due to the NP transport equation 
\begin{equation} \label{eq:transport1}
\overbar{\delta} n^{\mu} = \overbar{\mu} \mb^{\mu} - \left( \alpha + \overbar{\beta} \right) n^{\mu},
\end{equation} 
for the Kerr metric \cite{np62,pirani64,teu74}. 


Similar expressions can likewise be obtained for the other Ricci-NP quantities. The ones which we require are
\begin{equation} \label{eq:ricnp2}
\mb_{\mu} \mb_{\nu} R^{\mu \nu}_{h} = a_{2} \left[ \overbar{\delta} \overbar{\delta} - \left( \alpha - \overbar{\beta} \right) \overbar{\delta} \right] R_{h},
\end{equation}
\begin{equation} \label{eq:ricnp3}
n_{\mu} n_{\nu} R^{\mu \nu}_{h} = a_{2} \left[ \Delta \Delta + \left( \gamma + \overbar{\gamma} \right) \Delta \right] R_{h},
\end{equation}
\begin{equation} \label{eq:ricnp4}
\ell_{\mu} m_{\nu} R^{\mu \nu}_{h} = a_{2} \left[ \delta D - \left( \overbar{\alpha} + \beta \right) D + \overbar{\rho} \delta \right] R_{h},
\end{equation}
and
\begin{equation} \label{eq:ricnp5}
\ell_{\mu} \ell_{\nu} R^{\mu \nu}_{h} = DD R_{h}.
\end{equation}

\subsection{Perturbed Weyl scalars}

We now recall the procedure initially undertaken by Teukolsky to derive equations for the perturbed Weyl scalars $\boldsymbol{\psi}^{h}$ \cite{teu73,teu74}. Teukolsky considered non-vacuum perturbations to the Kerr metric and, only in the final step of the derivation, replaced terms involving the perturbed Ricci tensor by terms proportional to some potentially non-zero stress-energy tensor as per the Einstein equations. Using the form \eqref{eq:fofrfieldTEU} for the field equations, which appear the same as the Einstein equations though with some `effective' stress-energy tensor, we see that the procedure in $f(R)$ gravity is almost identical. In particular, the null tetrad contractions of the Ricci tensor which appear in the Teukolsky equations are precisely those given above in \eqref{eq:ricnp1}--\eqref{eq:ricnp5}. Of course, the right-hand side of the resulting Teukolsky equations containing the `source terms' will now contain terms proportional to the metric and its derivatives through $R_{h}$, and thus may not be immediately `solvable' in the same way as it is in GR (though see below).


We therefore find that, in $f(R)$ gravity, the Weyl scalar\footnote{As a consequence of Sachs' peeling theorem \cite{sachs}, the complex components of the Weyl scalar $\psi^{h}_{4}$ are related to the tensorial $+$ and $\times$ GW polarisation modes through $\psi_{4} = -\tfrac{1}{2} \tfrac {\partial^2} {\partial t^2} \left({h}_{+}  - i {h}_{\times}\right)$ \cite{teu73,schfofr} (though cf. Ref. \cite{hoffman13}). The GW energy flux is then evaluable through a theory-dependent gravitational energy-momentum pseudotensor \cite{willbook,enermom,suv18}. The relationships between other (massless) polarisation modes, if they exist in a theory, and the perturbed Weyl scalars are determined through the $E(2)$ classification of the theory \cite{gong18}.} $\psi^{h}_{4}$ satisfies the Teukolsky equation \cite{teu73}
\begin{widetext}
\begin{equation} \label{eq:teupsi4}
\begin{aligned}
&\hspace*{-0.48cm}\left[ \left( \Delta + 3 \gamma - \overbar{\gamma} + 4 \mu + \overbar{\mu} \right) \left( D + 4 \epsilon - \rho \right) - \left( \overbar{\delta} - \overbar{\tau} + \overbar{\beta} + 3 \alpha + 4 \pi \right) \left( \delta - \tau + 4 \beta \right) - 3 \psi^{g}_{2} \right] \psi^{h}_{4} =\\
&\hspace*{-0.48cm} \frac {a_{2}} {2} \left( \Delta + 3 \gamma - \overbar{\gamma} + 4 \mu + \overbar{\mu} \right) \left\{  \left( \Delta + 2 \gamma - 2 \overbar{\gamma} + \overbar{\mu} \right) \left[ \overbar{\delta} \overbar{\delta} - \left( \alpha - \overbar{\beta} \right) \overbar{\delta} \right] R_{h} - \left( \overbar{\delta} - 2 \overbar{\tau} + 2 \alpha \right) \left[ \overbar{\delta} \Delta - \overbar{\mu} \overbar{\delta} + \left( \alpha + \overbar{\beta} \right) \Delta \right] R_{h} \right\} + \\
&\hspace*{-0.48cm} \frac {a_{2}} {2} \left( \overbar{\delta} - \overbar{\tau} + \overbar{\beta} + 3 \alpha + 4 \pi \right) \left\{ \left( \overbar{\delta} - \overbar{\tau} + 2 \overbar{\beta} + 2 \alpha \right) \left[ \Delta \Delta + \left( \gamma + \overbar{\gamma} \right) \Delta \right] R_{h} - \left( \Delta + 2 \gamma + 2 \overbar{\mu} \right) \left[ \overbar{\delta} \Delta - \overbar{\mu} \overbar{\delta} + \left( \alpha + \overbar{\beta} \right) \Delta \right] R_{h} \right\},
\end{aligned}
\end{equation}
\end{widetext}
where $R_{h}$ is determined through \eqref{eq:perttrace} and we have explicitly used expressions \eqref{eq:ricnp1}--\eqref{eq:ricnp3}. A similar equation is satisfied by the scalar $\psi^{h}_{0}$ \cite{teu73},
\begin{widetext}
\begin{equation} \label{eq:teupsi0}
\begin{aligned}
&\left[ \left( D - 3 \epsilon + \overbar{\epsilon} - 4 \rho - \overbar{\rho} \right) \left( \Delta - 4 \gamma + \mu \right) - \left( \delta + \overbar{\pi} - \overbar{\alpha} - 3 \beta - 4 \tau \right) \left( \overbar{\delta} + \pi - 4 \alpha \right) - 3 \psi^{g}_{2} \right] \psi^{h}_{0} = \\
& -\frac {a_{2}} {2} \left( \delta + \overbar{\pi} - \overbar{\alpha} - 3 \beta - 4 \tau \right) \left\{ \left( D - 2 \epsilon - 2 \overbar{\rho} \right) \left[ \delta D - \left( \overbar{\alpha} + \beta \right) D + \overbar{\rho} \delta \right] R_{h} - \left( \delta +\overbar{\pi} - 2 \overbar{\alpha} - 2 \beta \right) DD R_{h} \right\} \\
& - \frac {a_{2}} {2} \left( D - 3 \epsilon + \overbar{\epsilon} - 4 \rho - \overbar{\rho} \right) \left\{ \left( \delta + 2 \overbar{\pi} - 2 \beta \right) \left[ \delta D - \left( \overbar{\alpha} + \beta \right) D + \overbar{\rho} \delta \right] R_{h} - \left( D - 2 \epsilon + 2 \overbar{\epsilon} - \overbar{\rho} \right) \left[ \delta \delta - \left( \overbar{\alpha} - \beta \right) \delta \right] R_{h} \right\},
\end{aligned}
\end{equation}
\end{widetext}
where we have used the conjugated version of \eqref{eq:ricnp2} together with \eqref{eq:ricnp4} and \eqref{eq:ricnp5}.

Equations \eqref{eq:teupsi4} and \eqref{eq:teupsi0} are to be subjected to boundary conditions which ensure that no radiation enters inward from infinity, and that no radiation escapes the black hole horizon \cite{teubcs}.

As noted before, independently of whether equations \eqref{eq:teupsi4} and \eqref{eq:teupsi0} can be readily integrated or not, they still provide a description for the perturbed Weyl scalars in $f(R)$ (or any other, provided the perturbed Ricci tensor is kept) theory of gravity. However, a critical feature of the $f(R)$ theory is the following: the `source' terms involving $R_{h}$ featured within \eqref{eq:teupsi4} and \eqref{eq:teupsi0} can be expressed in terms of \emph{background} quantities because the perturbed trace equation \eqref{eq:perttrace} completely determines $R_{h}$. As such, although the Teukolsky equations \eqref{eq:teupsi4} and \eqref{eq:teupsi0} contain metric pieces implicitly through $R_{h}$, they represent \emph{decoupled} differential equations for the Weyl scalars once a suitable solution to \eqref{eq:perttrace} has been selected.

Interestingly enough, however, upon expansion of the right-hand sides of \eqref{eq:teupsi4} and \eqref{eq:teupsi0}, one finds that all of these `source' terms involving $R_{h}$ cancel out exactly. Though we present it without proof because the algebra is long (though not especially difficult), employing the commutator properties \eqref{eq:com1}--\eqref{eq:com4} for the operators appearing on the right-hand sides of \eqref{eq:teupsi4} and \eqref{eq:teupsi0} shows that total cancellations occur, regardless of the functional form of $R_{h}$. In particular, the $f(R)$ Teukolsky equations for $\psi^{h}_{4}$ and $\psi^{h}_{0}$ over a Kerr background are identical to the vacuum GR case with vanishing right-hand side. This does not, however, imply that the metric perturbation is identical, as we show in the next section.


\section{Metric reconstruction}

As mentioned earlier, introducing a perturbation into the metric likewise introduces a perturbation into the null tetrads \eqref{eq:ellkerr}--\eqref{eq:mkerr}. In turn, perturbing the general expression \eqref{eq:kerrmet}, one can show that the tensorial components of the metric perturbation $\boldsymbol{h}$ take the form \cite{price07} (see also Appendix A)
\begin{equation} \label{eq:pertmetexp}
\begin{aligned}
h_{\mu \nu} =& h_{nn} \ell_{\mu} \ell_{\nu} - 2 h_{ n \mb} \ell_{ ( \mu} m_{ \nu )} - 2 h_{n m} \ell_{(\mu} \mb_{\nu )} \\
&+ 2 h_{\ell n} \ell_{( \mu} n_{\nu)} + h_{\ell \ell} n_{\mu} n_{\nu} - 2 h_{\ell \mb} n_{(\mu} m_{\nu)} \\
&- 2 h_{\ell m} n_{(\mu} \mb_{\nu)}  + h_{mm} \mb_{\mu} \mb_{\nu} \\
&+ 2 h_{m \mb} m_{(\mu} \mb_{\nu )} + h_{\mb \mb} m_{\mu} m_{\nu} ,
\end{aligned}
\end{equation}
where the round brackets denote the usual symmetrisation operation: $U_{(ij)k} \equiv \tfrac{1}{2} \left( U_{ijk} + U_{jik} \right)$. 

In vacuum GR, one can build all components of $\boldsymbol{h}$ in \eqref{eq:pertmetexp} through the introduction of generalised Debye potentials, which are related to solutions of the Teukolsky equations \eqref{eq:teupsi4} and \eqref{eq:teupsi0} \cite{chr75,perttet,coh75,otherkerrmet}. 

However, when sources are present or when modified gravity is being considered, the problem is more complicated. Specifically, the terms $\psi^{h}_{4}$ and $\psi^{h}_{0}$ alone are not sufficient to determine the metric for two reasons. The first is that the gauge conditions necessary for the construction of the generalised Debye potentials do not exist when the perturbed Einstein tensor does not satisfy certain symmetry properties \cite{price07}. In modified gravity, these symmetries are, in general, violated because additional, non-tensorial GW polarisation modes are excitable \cite{gwpols1,gwpols2}. In particular, imposing a transverse and traceless constraint on the perturbed metric \eqref{eq:pertmetexp} would \emph{a priori} forbid the possibility of these propagating modes \cite{starobinsky,massgws}. The second (though related) reason is that the Ricci identities (see below) contain terms proportional to $R_{h}^{\mu \nu}$ \cite{pirani64}, which need not vanish in vacuum for modified theories of gravity (see Sec. III A). These pieces cannot, in general, be expressed purely in terms of the Weyl scalars, though must enter into the metric \cite{ori03,barrack03}.


As in GR, we are free to exploit coordinate and tetrad freedoms to fix a gauge which simplifies the resulting algebra. 
In particular, owing to the fact that $\boldsymbol{\ell}$ is a repeated principal null direction (i.e. a repeated eigenvector of the Weyl tensor) for the Kerr metric \cite{petrov54}, we maintain enough freedom to set\footnote{In GR, one typically employs either the so-called  `in-going radiation' (ING) or `out-going radiation' (ORG) gauges initially developed by Chrzanowski \cite{chr75}. These gauges, however, incorporate a traceless condition on the perturbed metric, which is physically incompatible with the possibility of Ricci mode excitations \cite{gwpols1,gwpols2}. In general, as proven in Ref. \cite{price07}, when the perturbed Ricci tensor is not orthogonal to the null tetard $\bl$ ($\bn$), i.e. when $\ell_{\mu} \ell_{\nu} R_{h}^{\mu \nu}  \neq 0$ ($n_{\mu} n_{\nu} R^{\mu \nu} \neq 0$), the ING (ORG) does not exist [cf. expressions \eqref{eq:ricnp5} and \eqref{eq:ricnp3}, respectively]. Expression \eqref{eq:somegauge} ($\ell_{\mu} h^{\mu \nu} = 0$) is a weaker form of the ING.} \cite{bg11,massgws},
\begin{equation} \label{eq:somegauge}
h_{\ell \ell} = h_{\ell n} = h_{\ell m} = h_{\ell \mb} = 0 .
\end{equation}
Additionally, the simple observation that \cite{chrz76}
\begin{equation} \label{eq:conj1}
\overbar{h_{m m}} \equiv \overbar{ \left( m^{\mu} m^{\nu} h_{\mu \nu} \right)} = \mb^{\mu} \mb^{\nu} h_{\mu \nu} \equiv h_{ \mb \mb},
\end{equation}
and
\begin{equation} \label{eq:conj2}
\overbar{h_{n m}} \equiv \overbar{ \left( n^{\mu} m^{\nu} h_{\mu \nu} \right)} = n^{\mu} \mb^{\nu} h_{\mu \nu} \equiv h_{ n \mb},
\end{equation}
means that we need only determine four independent metric functions: $h_{nn}, h_{nm}, h_{m \mb}$, and $h_{m m}$. To achieve this, we require four independent equations. The Ricci identities fulfil this purpose, and the four equations we need, when expressed using the NP variables for an arbitrary spacetime, are given exactly by (e.g. page 350 of Pirani \cite{pirani64})
\begin{equation} \label{eq:ricid1}
\begin{aligned}
D \rho - \overbar{\delta} \kappa =& \rho^2 + \sigma \overbar{\sigma} + \left( \epsilon + \overbar{\epsilon} \right) \rho - \overbar{\kappa} \tau \\
&- \kappa \left( 3 \alpha + \overbar{\beta} - \pi \right) - \frac {1} {2} \ell_{\mu} \ell_{\nu} R^{\mu \nu},
\end{aligned}
\end{equation}
\begin{equation} \label{eq:ricid2}
\begin{aligned}
D \sigma - \delta \kappa =& \left( \rho + \overbar{\rho} \right) \sigma + \left( 3 \epsilon - \overbar{\epsilon} \right) \sigma \\
&- \left( \tau - \overbar{\pi} + \overbar{\alpha} + 3 \beta \right) \kappa + \psi_{0},
\end{aligned}
\end{equation}
\begin{equation} \label{eq:ricid3}
\begin{aligned}
D \alpha - \overbar{\delta} \epsilon =& \left( \rho + \overbar{\epsilon} - 2 \epsilon \right) \alpha + \beta \overbar{\sigma} - \overbar{\beta} \epsilon - \kappa \lambda \\
&- \overbar{\kappa} \gamma + \left( \epsilon + \rho \right) \pi - \frac {1} {2} \ell_{\mu} \mb_{\nu} R^{\mu \nu},
\end{aligned}
\end{equation}
and
\begin{equation} \label{eq:ricid4}
\begin{aligned}
\Delta \lambda - \overbar{\delta} \nu =& - \left( \mu + \overbar{\mu} + 3 \gamma - \overbar{\gamma} \right) \lambda \\
&+ \left( 3 \alpha + \overbar{\beta} + \pi - \overbar{\tau} \right) \nu - \psi_{4}.
\end{aligned}
\end{equation}


Perturbing expressions \eqref{eq:ricid1}--\eqref{eq:ricid4}, and making use of \eqref{eq:ricnp1}--\eqref{eq:ricnp5} together with \eqref{eq:somegauge} we find that, for a perturbed Kerr spacetime in $f(R)$ gravity,
\begin{equation} \label{eq:np1}
D \rho_{h} = 2 \rho \rho_{h} -  \frac {a_{2}} {2} DD R_{h},
\end{equation}
\begin{equation} \label{eq:np2}
D \sigma_{h} = \left( \rho + \overbar{\rho} \right) \sigma_{h} + \psi^{h}_{0},
\end{equation}
\begin{equation} \label{eq:np3}
\begin{aligned}
D \alpha_{h} - \overbar{\delta} \epsilon_{h} =& \rho \alpha_{h} + \rho_{h} \alpha + \beta \sigma_{h} - \overbar{\beta} \epsilon_{h} \\
&+ \left( \overbar{\epsilon}_{h} - 2 \epsilon_{h} \right) \alpha + \rho \pi_{h} + \left( \epsilon_{h} + \rho_{h} \right) \pi \\
&- \frac{a_{2}} {2} \left[ \overbar{\delta} D - \left( \alpha + \overbar{\beta} \right) D + \rho \overbar{\delta} \right] R_{h},
\end{aligned}
\end{equation}
and
\begin{equation} \label{eq:np4}
\begin{aligned}
 \overbar{\delta} \nu_{h} - \Delta \lambda_{h} =& \left( \mu + \overbar{\mu} + 3 \gamma - \overbar{\gamma} \right) \lambda_{h} \\
 &- \left(3 \alpha + \overbar{\beta} + \pi - \overbar{\tau} \right) \nu_{h} + \psi^{h}_{4},
 \end{aligned}
\end{equation}
respectively.

\subsection{Metric components}


In this section, we show that the set of equations \eqref{eq:np1}--\eqref{eq:np4} yields a linear, decoupled system which, when solved, allows for a total reconstruction of the metric \eqref{eq:pertmetexp}. Expanding equation \eqref{eq:np1} in terms of the components of $\boldsymbol{h}$ [using expression \eqref{eq:rhob} from Appendix A], we find that
\begin{equation} \label{eq:mete1}
\left( D - 2 \rho \right) \left( D + \rho - \overbar{\rho} \right) h_{m \mb} =  -a_{2} DD R_{h},
\end{equation}
where, again, we note that $R_{h}$ is determined through \eqref{eq:perttrace}, so that \eqref{eq:mete1} can be expressed purely in terms of background quantities. The differential operator appearing on the left-hand side of \eqref{eq:mete1} may be readily expressed in a coordinate basis using \eqref{eq:ellkerr} and \eqref{eq:kerrnpquant} together with the definition of the directional derivatives \eqref{eq:directionalderivatives}. Note that equation \eqref{eq:mete1} implies that one can set $h_{m \mb} = 0$ when $a_{2} =0$ without loss of generality, as in GR \cite{chr75,coh75,teubcs}. One can thus clearly see how the massive Ricci mode manifests in the perturbed metric.

From the second NP relationship \eqref{eq:np2}, we obtain a similar expression for $h_{mm}$, viz.
\begin{equation} \label{eq:mete2}
 \left( D - \rho - \overbar{\rho} \right) \left( D + \rho - \overbar{\rho} \right) h_{mm} = 2 \psi_{0}^{h},
\end{equation}
where $\psi_{0}^{h}$ is solved through \eqref{eq:teupsi0} and we have used \eqref{eq:sigmab}.

The expression for $h_{n m}=\overbar{h_{n\mb}}$ is slightly more complicated, though can still be expressed in terms of background terms and the other components given above. From expression \eqref{eq:np3}, using \eqref{eq:epsilonb}, \eqref{eq:pib}, and \eqref{eq:alphab} and grouping terms, we find
\begin{equation} \label{eq:mete3}
\begin{aligned}
&\left[ \left( D - \rho \right) \left( D - 2 \overbar{\rho} - \rho \right) + 2 \rho \left( D - \rho \right) \right] h_{n \mb} = \\
& \Big[\overbar{\delta} \left( \rho - \overbar{\rho} \right)  - \left( D - \rho \right) \left( \overbar{\delta} + 2 \alpha - \pi - \overbar{\tau} \right) \\
&+ \alpha \left(2 D + \overbar{\rho} - \rho \right)  - \overbar{\beta} \left( \rho - \overbar{\rho} \right) \\
&- 2 \rho \overbar{\tau} + \pi \left(2D + 3 \rho - 3 \overbar{\rho} \right) \Big] h_{m \mb}\\
&+ \left[ \left(D - \rho \right) \left( \delta - 2 \overbar{\alpha} + \overbar{\pi} + \tau \right) - 2 \rho \tau \right] h_{\mb \mb} \\
&+ 2 \beta \left( D + \rho - \overbar{\rho} \right) h_{m m} \\
&- 2 a_{2} \left[ \overbar{\delta} D - \left( \alpha + \overbar{\beta} \right) D + \rho \overbar{\delta} \right] R_{h}  ,
\end{aligned}
\end{equation}
where $h_{m \mb}$ is given through \eqref{eq:mete1} and $h_{m m} = \overbar{h_{ \mb \mb}}$ is given through \eqref{eq:mete2}. 

Finally, we determine $h_{n n}$ through expression \eqref{eq:np4},
\begin{equation} \label{eq:mete4}
\begin{aligned}
&\hspace*{-0.52cm} \left( \overbar{\delta} + 3 \alpha + \overbar{\beta} + \pi - \overbar{\tau} \right) \left( \overbar{\delta} + 2 \alpha + 2 \overbar{\beta} - \pi - \overbar{\tau} \right) h_{n n} = \\
&\hspace*{-0.45cm} 2 \Big[ \left( \overbar{\delta} + 3 \alpha + \overbar{\beta} + \pi - \overbar{\tau} \right) \left( \Delta + \overbar{\mu}  + 2 \gamma \right)  \\
&\hspace*{-0.52cm} - \left( \Delta + \mu + \overbar{\mu} + 3 \gamma - \overbar{\gamma} \right)  \left( \overbar{\tau} + \pi \right) \Big] h_{n \mb} + 2 \psi_{4}^{h} \\
&\hspace*{-0.52cm} - \left( \Delta + \mu + \overbar{\mu} + 3 \gamma - \overbar{\gamma} \right) \left( \Delta + \overbar{\mu} - \mu + 2 \gamma - 2 \overbar{\gamma} \right) h_{\mb \mb} ,\\
\end{aligned}
\end{equation}
where one uses \eqref{eq:nub}, \eqref{eq:lambdah}, \eqref{eq:teupsi4}, and evaluates $h_{n \mb}$ via \eqref{eq:mete3}. 

Therefore, keeping in mind the conjugations \eqref{eq:conj1} and \eqref{eq:conj2} and the gauge choice \eqref{eq:somegauge}, we have shown that expressions \eqref{eq:mete1}--\eqref{eq:mete4} completely describe the geometry surrounding a gravitationally perturbed Kerr black hole in $f(R)$ gravity. Importantly, these equations are all decoupled, and thus can be solved using standard techniques suited to linear, second order partial differential equations. Just as in GR, the metric equations \eqref{eq:mete1}--\eqref{eq:mete4} are subject to boundary conditions which prevent radiation entering from infinity or escaping from the black hole \cite{teubcs}. An explicit solution to these equations will be presented elsewhere. 



\section{Discussion}

In GR, the geometry surrounding an isolated black hole must be Kerr \cite{cart1,rob1,mars1}. Searches for non-Kerr features in astrophysically stable black holes therefore provide one of the best means to probe GR in the strong field regime \cite{bhtests,gurl15,papa1,bambi08,gw17}. In contrast, validations of the Kerr metric do not necessarily signal that GR alone describes the gravitational field of black holes because the Kerr solution is common to many theories of gravity \cite{kerrinmod}. However, the way in which a black hole responds to disturbances depends intimately on the structure of the field equations \cite{pertkerrfr,morekerr}. Astrophysically distorted Kerr black holes can therefore still provide a natural laboratory to probe GR. Though theory-dependent signatures are sub-leading in this case, they may be detectable with upcoming precision GW experiments \cite{subtle}. In this paper, we have provided tools which allow for a simple description of gravitational perturbations of the Kerr spacetime in the $f(R)$ family of theories. The major results are encoded within equations \eqref{eq:mete1}--\eqref{eq:mete4}, which provide a set of linear, decoupled equations for the perturbed metric tensor components. The dependence of the perturbed fields on the theory is evident through the appearance of the quadratic $f(R)$ parameter $a_{2}$ defined in \eqref{eq:maclaurin}. Though we have focused on the Kerr metric, our results could easily be extended to any type II Einstein background \cite{petrov54}.

Under favourable conditions, a GW incident on a rapidly rotating Kerr black hole in GR can have its amplitude \emph{increased} by up to $\approx 138 \%$ \cite{teu74}. This effect, known as superradiance, can lead to an instability (coined the `black-hole bomb') wherein trapped waves are repeatedly amplified \cite{bhbomb}. This scenario potentially has astrophysical consequences, such as for the production of relativistic jets \cite{astrocon,astrocon2}.  The results presented here could be a starting point for one to investigate gravitational superradiance in $f(R)$ gravity, and, more generally, GW scattering by black holes in $f(R)$ gravity. Incidentally, massive scalar fields are known to be particularly prone to superradiant instabilities \cite{masskg}. This fact may have important consequences for $f(R)$ gravity since the Ricci mode satisfies the Klein-Gordon equation \eqref{eq:perttrace} \cite{myung13} (see also Refs. \cite{card13,card14}). 

Massive modes can be searched for directly using facilities such as the Laser Interferometer Gravitational-Wave Observatory (LIGO) \cite{testspol,testspol2}. Moreover, the data from GW170104 place an upper bound on the graviton mass $m_{\text{graviton}} < 7.7 \times 10^{-23} \text{ eV}/c^{2}$ \cite{gw17}, which applies directly to the parameter $a_{2}$ within \eqref{eq:maclaurin}. In general, following a compact object merger or core collapse event, the remnant object is expected to be born into a ringdown phase wherein it oscillates (e.g. \cite{kokqnm}). An analysis of the associated quasi-normal mode features may then further constrain $a_{2}$ \cite{qnm2,qnm3}. Moreover, in treatments of the two-body problem (especially relevant to the early stages of an inspiral), an explicit form for the metric is required to model self-force effects \cite{ori03,barrack03}. The results presented in this paper may then prove useful in a study of black hole back-reaction in an $f(R)$ theory.



\section*{Acknowledgements}
We thank Prof. Kostas Kokkotas for discussions. We thank the anonymous referee for their helpful suggestions. This work was supported by the Alexander von Humboldt Foundation.


\appendix

\section{Newman-Penrose expressions}

Here we present expressions for the NP spin-coefficients ($\kappa, \rho, \sigma, \ldots$) in terms of the components of the perturbed metric tensor $h_{\mu \nu}$. In general, the freedom offered through infinitesimal tetrad transformations allows for the perturbed null tetrad to be expressed as (see expressions (2.1)--(2.6) of Ref. \cite{perttet})

\begin{equation} \label{eq:pertl}
\ell^{\mu}_{h} = - \frac {1} {2} h_{\ell \ell} n^{\mu},
\end{equation}
\begin{equation} \label{eq:pertn}
n^{\mu}_{h} = - \frac {1} {2} h_{nn} \ell^{\mu} - h_{n \ell} n^{\mu},
\end{equation}
and
\begin{equation} \label{eq:pertm}
m^{\mu}_{h} = \frac {1} {2} h_{mm} \mb^{\mu} + \frac {1} {2} h_{m \mb} m^{\mu} - h_{m \ell} n^{\mu} - h_{m n} \ell^{\mu}.
\end{equation}
From expressions \eqref{eq:pertl}--\eqref{eq:pertm} one can readily define the perturbed NP derivatives \eqref{eq:directionalderivatives} and their commutation relations. In general, these latter relations read (e.g. \cite{pirani64})
\begin{equation} \label{eq:com1}
\Delta D - D \Delta = \left( \gamma + \overbar{\gamma} \right) D + \left( \epsilon + \overbar{\epsilon} \right) \Delta - \left( \overbar{\tau} + \pi \right) \delta - \left( \tau + \overbar{\pi} \right) \overbar{\delta},
\end{equation}
\begin{equation} \label{eq:com2}
\delta D - D \delta = \left( \overbar{\alpha} + \beta - \overbar{\pi} \right) D + \kappa \Delta - \left( \overbar{\rho} + \epsilon - \overbar{\epsilon} \right) \delta - \sigma \overbar{\delta},
\end{equation}
\begin{equation} \label{eq:com3}
\delta \Delta - \Delta \delta = - \overbar{\nu} D + \left( \tau - \overbar{\alpha} - \beta \right) \Delta + \left( \mu + \overbar{\gamma} - \gamma \right) \delta - \overbar{\lambda} \overbar{\delta},
\end{equation}
and
\begin{equation} \label{eq:com4}
\overbar{\delta} \delta - \delta \overbar{\delta} = \left( \overbar{\mu} - \mu \right) D + \left( \overbar{\rho} - \rho \right) \Delta + \left( \alpha - \overbar{\beta} \right) \delta + \left( \beta - \overbar{\alpha} \right) \overbar{\delta}.
\end{equation}
Expressions \eqref{eq:pertl}--\eqref{eq:pertm} allow for the NP spin-coefficients to be expressed in terms of the metric components by matching the coefficients in expressions \eqref{eq:com1}--\eqref{eq:com4}. The spin-coefficients are then given through (see equations (A4) of Ref. \cite{lousto99})
\begin{equation}
\kappa_{h} = \left( D - \overbar{\rho} - 2 \epsilon \right) h_{\ell m} - \frac {1} {2} \left( \delta - 2 \alpha - 2 \beta + \overbar{\pi} + \tau \right) h_{\ell \ell},
\end{equation}
\begin{equation} \label{eq:sigmab}
\sigma_{h} = \left( \overbar{\pi} + \tau \right) h_{\ell m} + \frac {1} {2} \left( D + \rho - \overbar{\rho} + 2 \overbar{\epsilon} - 2 \epsilon \right) h_{mm} ,
\end{equation}
\begin{equation} \label{eq:nub}
\nu_{h} = - \left( \Delta + \overbar{\mu} + 2 \gamma \right) h_{n \mb} + \frac {1} {2} \left( \overbar{\delta} + 2 \alpha + 2 \overbar{\beta} - \pi - \overbar{\tau} \right) h_{nn},
\end{equation}
\begin{equation} \label{eq:lambdah}
\lambda_{h} = - \left( \overbar{\tau} + \pi \right) h_{n \mb} - \frac {1} {2} \left( \Delta + \overbar{\mu} - \mu + 2 \gamma - 2 \overbar{\gamma} \right) h_{\mb \mb} ,
\end{equation}
\begin{equation}
\begin{aligned}
2 \mu_{h} =& \rho h_{nn} - \left( \delta + 2 \beta + \tau \right) h_{n \mb} + \left( \overbar{\delta} + 2 \overbar{\beta} - 2 \pi - \overbar{\tau} \right) h_{nm} \\
&- \frac {1} {2} \left( 2 \Delta + \overbar{\mu} - \mu + \gamma - \overbar{\gamma} \right) h_{m \mb} ,
\end{aligned}
\end{equation}
\begin{equation} \label{eq:rhob}
\begin{aligned}
 2 \rho_{h} =& \overbar{\mu} h_{\ell \ell} + \left( \rho - \overbar{\rho} \right) h_{n \ell} + \left( D + \rho - \overbar{\rho} \right) h_{m \mb} \\
 &- \left( \delta - 2 \overbar{\alpha} - \overbar{\pi} \right) h_{ \ell \mb} + \left( \overbar{\delta} + 2 \overbar{\tau} - 2 \alpha + \pi \right) h_{\ell m},
 \end{aligned}
\end{equation}
\begin{equation} \label{eq:epsilonb}
\begin{aligned}
 2 \epsilon_{h}  =& \left( D + \rho - \overbar{\rho} \right) h_{n \ell} + \frac {1} {2} \left( \overbar{\delta} - 2 \alpha - \pi \right) h_{\ell m} \\
& - \frac {1} {2} \left( \delta - 2 \overbar{\alpha} +3 \pi + 4 \tau \right) h_{\ell \mb}  \\
&+ \frac {1} {2} \left( \rho - \overbar{\rho} \right) h_{m \mb} - \frac {1} {2} \left( \Delta + 2 \gamma \right) h_{\ell \ell},
\end{aligned}
\end{equation}
\begin{equation} \label{eq:pib}
\begin{aligned}
2 \pi_{h} =& - \left( D - \rho -2 \epsilon \right) h_{n \mb} - \left( \overbar{\delta} + \overbar{\tau} + \pi \right) h_{n \ell} \\
&- \left( \Delta + \overbar{\mu} - 2 \overbar{\gamma} \right) h_{\ell \mb} - \overbar{\tau} h_{m \mb} - \tau h_{\mb \mb} ,
\end{aligned}
\end{equation}
\begin{equation}
\begin{aligned}
2 \tau_{h} =& \left( D - \overbar{\rho} + 2 \overbar{\epsilon} \right) h_{nm} + \left( \delta - \overbar{\pi} - \tau \right) h_{n \ell} \\
&+ \left( \Delta + \mu - 2 \gamma \right) h_{\ell m} - \overbar{\pi} h_{m \mb} - \pi h_{ m m},
\end{aligned}
\end{equation}
\begin{equation} \label{eq:alphab}
\begin{aligned}
4 \alpha_{h} =&  \left( D - 2 \overbar{\rho} - \rho - 2 \epsilon \right) h_{n \mb} \\
&- \left( \Delta + 4 \gamma - 2 \mu + \overbar{\mu} - 2 \overbar{\gamma} \right) h_{\ell \mb} \\
&-  \left( \overbar{\delta} + \pi + \overbar{\tau} \right) h_{n \ell} \\
&+  \left( \overbar{\delta} + 2 \alpha - \pi - \overbar{\tau} \right) h_{ m \mb} \\
&- \left( \delta - 2 \overbar{\alpha} + \overbar{\pi} + \tau \right) h_{\mb \mb},
\end{aligned}
\end{equation}

\begin{equation}
\begin{aligned}
4 \beta_{h} =& \left( D - \overbar{\rho} - 4 \epsilon + 2 \rho + 2 \overbar{\epsilon} \right) h_{nm } \\
&-  \left( \Delta + \mu + 2 \overbar{\mu} + 2 \gamma \right) h_{\ell m} \\
&- \left( \delta + \overbar{\pi} + \tau \right) h_{n \ell} \\
&-  \left( \delta - 2 \beta + \overbar{\pi} + \tau \right) h_{ m \mb} \\
&+  \left( \overbar{\delta} + 2 \beta - \pi - \overbar{\tau} \right) h_{m m},
\end{aligned}
\end{equation}
and finally
\begin{equation}
\begin{aligned}
2 \gamma_{h} =& - \left( \overbar{\gamma} + \gamma \right) h_{n \ell} + \frac {1} {2} \left( D + \rho - \overbar{\rho} + 2 \overbar{\epsilon} \right) h_{nn} \\
&- \frac {1} {2} \left( \delta + 2 \beta + 2 \overbar{\pi} + 3 \tau \right) h_{n \mb} \\
&+ \frac {1} {2} \left( \overbar{\delta} + 2 \overbar{\beta} - 2 \pi - \overbar{\tau} \right) h_{n m} \\
&+ \frac {1} {4} \left( 3 \overbar{\mu} - 2 \mu + \gamma - \overbar{\gamma} \right) h_{m \mb}.
\end{aligned}
\end{equation}





\end{document}